\documentstyle[12pt]{article}

\newcommand{\vs}[1]{\vspace{#1 mm}}

\newcommand{\beq}{\begin{equation}}
\newcommand{\eeq}{\end{equation}}
\newcommand{\beqa}{\begin{eqnarray}}
\newcommand{\eeqa}{\end{eqnarray}}
\newcommand{\nn}{\nonumber}
\newcommand{\eq}[1]{(\ref{#1})}

\newcommand{\su}{{\cal U}_q (sl(2))}
\newcommand{\Db}{\bar \Delta}
\begin{document}
\topmargin 0pt
\oddsidemargin 5mm
\begin{titlepage}
\setcounter{page}{0}
\begin{flushright}
 OS-GE-39-93 \\
 October 1993
\end{flushright}
\vs{15}
\begin{center}
{\large Quantum Group Symmetry and Quantum Hall \\
Wavefunctions on a Torus}
\vs{20}

Haru-Tada Sato$^\dagger$
\vs{8}

{\em Institute of Physics, College of General Education \\
Osaka University, Toyonaka, Osaka 560, Japan}
\end{center}
\vs{10}

\begin{abstract}
We find a quantum group structure in two-dimensional motion
of nonrelativistic electrons in a uniform magnetic field on a torus.
The representation basis of the quantum algebra is composed of the
quantum Hall wavefunctions proposed by Haldane-Rezayi at the
Landau-level filling factor $\nu=1/m$ ($m$ odd). It is also shown
that the quantum group symmetry is relevant to the degenerate Landau
states and the deformation parameter of the quantum algebra is given by
the filling factor.
\end{abstract}
\vspace{1cm}

%
\noindent-----------------------------------------------------------------------
$^{\dagger}$ {\footnotesize Fellow of the Japan Society for the
Promotion of Science} \\
\mbox{}\hspace{0.4cm}{\footnotesize E-mail address~: hsato@jpnyitp.
yukawa.kyoto-u.ac.jp}
\end{titlepage}
\newpage
%
%

There have been many discussions in the study of quantum groups
(quantum universal enveloping algebras \cite{DJ,KS}). Quantum group
structures are found in (2+1)-dimensional topological Chern-Simons
theories \cite{CS} as well as in rational conformal field theories and
integrable lattice models \cite{YB}. Although the abelian Chern-Simons
theory does not possess a quantum group structure in the literature
\cite{CS}, it might be possible to exhibit one in some other senses.
There have been also interesting investigations of condensed matter
problems such as the fractional quantum Hall effect \cite{QH} making use of
the abelian Chern-Simons theory.

We have considered the two-dimensional planer motion of a
nonrelativistic particle in a uniform magnetic field in the
previous paper \cite{sato} and we derived the quantum group algebra $\su$
acting within each Landau level. It can be understood from the result
of that paper that the wavefunctions form a representation basis of the
$\su$ algebra and that the deformation parameter $q$ is related to the
filling factor $\nu=1/(2j+1)$ by $q=exp(2\pi i\nu)$. These observations
motivate us to discuss a relation between quantum groups and the
quantum Hall effects.

In this report we consider a $N_e$-electron system in the lowest
Landau level on a square torus of side $L$. We show that this case also
exhibits a quantum group symmetry similarly as the planer case and the
deformation parameter $q$ is given by the filling factor $\nu=1/m$
($m$ odd integer).

We use the results of the previous paper and thus start with reviewing
some formulae. In the system of a charged particle in a constant
magnetic field $B$ perpendicular to the $x$-$y$ plane, the generators
of the quantum group algebra are realized by the magnetic translation
operators $T_{\alpha}$ \cite{mag}
\beq
T_{(\alpha_1,\alpha_2)}=
exp({i\over\hbar}{\bf \alpha}\cdot{\bf \beta})\,,\label{EA}\eeq
where
\beq
\beta_i = p_i -{e\over c}A_i - {eB\over c}\epsilon_{ij}x^j\,,\label{eq2}
\eeq
\beq
\epsilon_{11}=\epsilon_{22}=0,\hskip 25pt
\epsilon_{12}=-\epsilon_{21}=1, \label{eq3}\eeq
and the gauge potential is
\beq
A_i = -{1\over2}B\epsilon_{ij}x^j -\partial_i\Lambda.\label{eq4}\eeq
The vector $\beta$ is related to the cyclotron center and we choose the
scalar function $\Lambda$ to be $\Lambda={1\over2}Bxy$ for simplicity.
The following combinations of magnetic translations \cite{sato}
\beq
E^{+}={T_{(\Delta, \Db)}-T_{(-\Delta, \Db)}
             \over q-q^{-1}} \label{EB} \eeq
\beq
E^{-}={T_{(-\Delta, -\Db)}-T_{(\Delta, -\Db)}
              \over q-q^{-1}} \label{EC} \eeq
\beq
k=T_{(\Delta,0)}. \label{ED} \eeq
satisfy the defining relations of the $\su$
\beq
[E^{+},E^{-}]
={ k^2 - k^{-2} \over q-q^{-1}}\,\,, \hskip 30pt
  k E^{\pm} k^{-1}=q^{\pm1}E^{\pm}\label{EE}\eeq
with the identification
\beq
 q=exp(i\Delta\Db l_B^{-2})\,. \label{EF} \eeq
The quantity $l_B$ is called a magnetic length $l_B=\sqrt{\hbar c/eB}$.
If we put $\Delta=L_x/(2j+1)$ and $\Db=L_y/(2j+1)$, the Landau states
$\psi_l$ ($-j\leq l\leq j$) \cite{LL} behave as a spin-$j$ representation
of $\su$. Namely, we obtain the relatons \cite{sato}
\beq
E^{\pm}\psi_l=[{1\over2}\pm l]_q\psi_{l\pm1}\,,\label{EG} \eeq
\beq
k\psi_l=q^l\psi_l\,,\label{EH} \eeq
where the notation $[x]_q$ means
\beq[x]_q={q^x-q^{-x}\over q-q^{-1}}. \label{eq12}\eeq
We note that $k$ measures the quantum number $l$ and $E^{+}$ ($E^{-}$)
raises (lowers) the $l$ and that our quantum algebra is associated with
only the quantum number $l$, namely the degeneracy of the Landau
levels. This means that the energy level (the Landau level) is
invariant under the action of the quantum algebra. This is the
difference from the case of the $su(2)$ angular momentum algebra.

Now we generalize the above argument to the case of a torus. The
quantum Hall wavefunctions on a torus are proposed by Haldane and Rezayi
\cite{HR} and discussed in some other papers \cite{La,Cr}. Our aim is to
examine whether the similar relation as \eq{EG} and \eq{EH} holds when we
operate the generators \eq{EB}-\eq{ED} to the wavefunction basis. First, we
consider the one-particle system on a torus in the lowest Landau level.
For simplicity, we specify the case of a square torus of side $L$ in
the unit $l_B=1$ and the gauge $\Lambda={1\over2}Bxy$ \cite{La,Cri}. The
lowest Landau-level wavefunction is written as
\beq
\psi=exp(-{1\over2}y^2)f(z) \label{eq13}\eeq
where $f(z)$ is an analytic function of $z=x+iy$ which satisfies the
periodicity condition
\beq
f(z)=e^{-{1\over2}L^2}e^{iLz}f(z+iL)\,.\label{EI}\eeq
If the total flux $N_s$ through the surface of the torus satisfies the
condition
\beq L^2 = 2\pi N_s\,,\label{EJ}\eeq
we have $N_s$ linearly independent solutions of \eq{EI} \cite{La}
\beq
f_l(z)=\Theta\left[\matrix
{{l\over N_s}\cr 0\cr}\right]({N_s\over L}z\vert iN_s)\,,\hskip 20pt
l=1,\dots,N_s \label{eq16}\eeq
where
\beq
\Theta\left[\matrix{a\cr b\cr}\right](w\vert\tau)=\sum_{n\in{\bf Z}}
exp\{i\pi\tau(n+a)^2+2\pi i(n+a)(w+b)\}\,.
\label{eq17}\eeq
For the convenience of notations, let us introduce the following
magnetic translations \cite{Cri}
\beq
S_{(n,m)}\equiv T_{(an,am)}\,,\hskip 30pt a={L\over N_s}\,,\label{eq18}
\eeq
With the use of \eq{EJ}, we can verify that these operators and the
wavefuctions
\beq
\psi_l=exp(-{1\over2}y^2)f_l(z) \label{eq19}\eeq
satisfy the following relations
\beqa
&S_{(n,0)}\psi_l(z)=e^{2\pi i{l\over N_s}n}\psi_l(z)\,,\nn \\
&S_{(0,n)}\psi_l(z)=\psi_{l+n}(z)\,, \label{EK}\eeqa
and the multiplication law
\beq
S_{(n,m)}=e^{-{i\over2}nma^2}S_{(n,0)}S_{(0,m)}\,.\label{EL}\eeq
Combining \eq{EK}, \eq{EL} and \eq{EJ}, we get
\beq
S_{(n,m)}\psi_l(z)=exp\left[2\pi i{n\over N_s}({m\over2}+l)\right]
\psi_{l+n}(z)\,.\label{EM}\eeq
Putting $\Delta=an$, $\Db=am$ in \eq{EB}-\eq{EF}, we easily recognize that
\beq
E^{+}={S_{(n,m)}-S_{(-n,m)}\over q-q^{-1}}\,,\hskip 15pt
E^{-}={S_{(-n, -m)}-S_{(n, -m)}\over q-q^{-1}}\,,\hskip 15pt
k=S_{(n,0)} \label{EN} \eeq
satisfy the $\su$ algebra \eq{EE} with
\beq
q=exp(ia^2nm)\,.\label{EO}\eeq
We thus obtain the action of the quantum algebra on the torus
\beqa
&E^{\pm}\psi_l(z)={
exp\left[2\pi i{n\over N_s}({m\over2}\pm l)\right]-
exp\left[-2\pi i{n\over N_s}({m\over2}\pm l)\right]
\over q-q^{-1}} \psi_{l+m}(z)\,,\nn \\
&k\psi_l(z)=exp(2\pi i {n\over N_s}l)\,.  \label{EP}\eeqa
We consequently have the same equations as \eq{EG} and \eq{EH} when we
put $n=m=1$ in the above equations \eq{EP}. It can be seen from \eq{EO}
that the value of $q$ is related to the filling factor
\beq
\nu \equiv {N_e \over N_s } \label{eq26} \eeq
by the relation
\beq
q=exp(2\pi i\nu)\,.\label{EQ}\eeq
The above case corresponds to $N_e=1$ and the other cases ($N_e\not=1$)
are discussed below.

Next we discuss the case of $N_e$-electron wavefunctions. Needless
to say in the non-interacting particle system, the comultiplication law
of the quantum algebra gives rise to the quantum group generators on
the tensor product representations of each particle's wavefunction;
\beq
E^{\pm}=\sum_{i=1}^{N_e}k_1 \dots k_{i-1} E_i^{\pm} k_{i+1}^{-1}
        \dots k_{N_e}^{-1}\,,\hskip 25pt k=\prod_{i=1}^{N_e}k_i
\eeq
and
\beq
\psi_{l_1,\dots,l_{N_e}}=\otimes_{i=1}^{N_e}\psi_{l_i}(x_i,y_i)\,.
\eeq
On the other hand, our intereset being a system with the filling
factor given by
\beq
\nu={1\over m}\,\hskip 30pt
    (\,m \hskip 10pt\rm{odd}\hskip 5pt\rm{integer}\,)\,, \label{eq28}
\eeq
we work with the wavefunctions of interacting particle system on the
torus in Ref.\cite{Cri}
\beq
\psi_l=exp[-{1\over2}\sum_{i=1}^{N_e}y_i^2]f_l(z_1,\dots,z_{N_e})\,,
\label{Ea}\eeq
\beq
f_l(z_1,\dots,z_{N_e})=\prod_{j<k}^{N_e}\left[{\Theta_1(z_j-z_k\vert i)
\over\Theta_1(0\vert i)}\right]^m \Theta\left[\matrix
{{l\over m}\cr 0\cr}\right]({m\over L}\sum_j z_j\vert im)\,, \label{eq30}
\eeq
where the function $\Theta_1$ is the elliptic theta function. Similarly
to the above case of the one-particle system, we introduce the magnetic
translation operators for the multi-particle system and operate them on
the wavefunctions \eq{Ea}. The total magnetic translation operator is
defined by the products of the $N_e$ copies of the one-particle magnetic
translation
\beq
S^{tot}_{(n,m)}=\prod_{i=1}^{N_e}S^{(i)}_{(n,m)}(x_i,y_i).
\label{eq31}\eeq
We then have the following relations instead of \eq{EK}
\beqa
&S^{tot}_{(n,0)}\psi_l(z)=e^{2\pi i{N_e\over N_s}ln}\psi_l(z)\,,\nn\\
&S^{tot}_{(0,n)}\psi_l(z)=\psi_{l+n}(z)\,,\label{ER}\eeqa
and the multiplication formula for $S^{tot}$ owing to \eq{EL}
\beq
S^{tot}_{(n_1,n_2)}S^{tot}_{(m_1,m_2)}
=exp\left({i\over2}a^2N_e\epsilon^{ij}n_im_j\right)
S^{tot}_{(n_1+n_2,m_1+m_2)}\,.\label{ES}\eeq
The generators of the quantum group algebra are obtained by replacing
$S$ with $S^{tot}$ in \eq{EN} and the commutation relations \eq{EE} are
checked using the above formula \eq{ES} when we set
\beq
q=exp(ia^2N_e nm)\,.\label{ET}\eeq
We thus find the similar equations as \eq{EP}
\beqa
&E^{\pm}\psi_l(z)={
exp\left[2\pi i\nu n({m\over2}\pm l)\right]-
exp\left[-2\pi i\nu n({m\over2}\pm l)\right]
\over q-q^{-1}} \psi_{l+m}(z)\,,\nn\\
&k\psi_l(z)=exp(2\pi i\nu l)\,.\label{EU}\eeqa
Again putting $n=m=1$, we verify that \eq{ET} coincides with \eq{EQ} and
the equations \eq{EU} become the same relations as \eq{EG} and \eq{EH}.

In this letter we showed that the quantum Hall wavefunctions on a torus
form a representation basis of the quantum algebra $\su$ of which
deformation parameter $q$ is related to the filling factor $\nu=1/m$
($m$ odd integer) by the relation $q=exp(2\pi i\nu)$, i.e., \eq{EQ}.
This statement is supported also by the result of Ref.\cite{sato} which
referred to the planer one-particle case. It is interesting to note that
the relation between $q$ and $\nu$ appears in accordance with $q$ being
a $m$-th root of unity. We have discussed only the case of a
primitive $m$-th root of unity, however we should investigate into
other $m$-th root cases in order to completely understand the relation
\eq{EQ} and the relevance of the quantum group structure to the quantum
Hall systems. Furthermore it is speculated that the representation
theory of quantum groups is expected to give a new approach to the
quantum Hall effects as well as to some other topics of the
two-dimensional electrons such as anyon systems \cite{any}.

\vspace{1cm}
\noindent
{\em Acknowledgments}

The author would like to thank N. Aizawa for a useful suggestion.
\newpage
%

%
%

\begin{thebibliography}{99}
%
\bibitem{DJ} V. Drinfeld, Proc. ICM-86, Berkeley, 1986, pp.798-820;\\
   M. Jimbo, Lett. Math. Phys. {\bf 10} (1985) 63; {\bf 11} (1986) 247.
\bibitem{KS} P. P. Kulish and N. Yu. Reshetikhin, J. Sov. Math. {\bf 23} (1983)
      \\ E. K. Sklyanin, Usp. Math. Nauk. {\bf 40} (1985) 214.
\bibitem{CS} E. Witten, Commun. Math. Phys. {\bf 121} (1989) 351;\\
       G. Siopsis, Mod. Phys. Lett. {\bf A6} (1991) 1515;\\
       E. Guadagnini, M. Martellini and M. Mintchev, Nucl. Phys.
       {\bf B336} (1990) 581.
\bibitem{YB} "Yang-Baxter Equation in Integrable Systems", ed. M. Jimbo
       (World Scientific, Singapore, 1990).
\bibitem{QH} "The Quantum Hall Effect", ed. R. E. Prange and S. M. Girvin
       (Springer-Verlag, New York, 1986).
\bibitem{LL} L. Landau and E. Lifshitz, "Quantum Mechanics", 3rd ed.
       (Pergamon, Oxford, 1977).
\bibitem{mag} E. Brown, Phys. Rev. {\bf 133} (1964) A1038;\\
         J. Zak, Phys. Rev. {\bf 134} (1964) A1602; ibid. A1607.
\bibitem{HR} F.D.M. Haldane and E.H. Rezayi, Phys. Rev. {\bf B31} (1985) 2529.
\bibitem{La} R.B. Laughlin, Ann. Phys. {\bf 191} (1989) 163.
\bibitem{Cr} G. Cristofano, G. Maiella, R. Musto and F. Nicodemi, Phys. Lett.
       {\bf B262} (1991) 88.
\bibitem{Cri} G. Cristofano, G. Maiella, R. Musto and F. Nicodemi,
        Mod. Phys. Lett. {\bf A6} (1991) 1779; ibid. 2985.
\bibitem{sato} H.-T. Sato, "Landau Levels and Quantum Group", Osaka Univ.
         preprint OS-GE-36-93.
\bibitem{any} A. Lerda and S. Sciuto, Nucl. Phys. {\bf B401} (1993) 613;
        \\ R. Caracciolo and M.A.R. Monteiro, Phys. Lett.
        {\bf B308} (1993) 58.
%
\end{thebibliography}
\end{document}